\documentclass[journal]{IEEEtran}

\usepackage{cite}
\usepackage{amsmath,amssymb,amsfonts}
\usepackage{graphicx}
\usepackage{textcomp}
\usepackage{float}
\usepackage{multirow}
\usepackage{siunitx} 
\usepackage[ruled,vlined,linesnumbered]{algorithm2e}
\usepackage{listings}
\usepackage{tabularx}
\usepackage{flushend}

\def\BibTeX{{\rm B\kern-.05em{\sc i\kern-.025em b}\kern-.08em
		T\kern-.1667em\lower.7ex\hbox{E}\kern-.125emX}}

\DeclareSIUnit{\GMAC}{GMAC}
\DeclareSIUnit{\FPS}{FPS}
\DeclareSIUnit{\GOPS}{GOPS}
\DeclareSIUnit{\bit}{b}
\DeclareSIUnit\permille{\text{\textperthousand}}

\begin{document}

\title{BinArray: A Scalable Hardware Accelerator for
	Binary Approximated CNNs}
\author{\IEEEauthorblockN{Mario Fischer, Juergen Wassner}\\
	\IEEEauthorblockA{\textit{Department of Engineering and Architecture, Lucerne University of Applied Sciences and Arts, Switzerland.}\\ juergen.wassner@hslu.ch}}

\markboth{Lucerne University of Applied Sciences and Arts, December~2020}{}
\maketitle
\thispagestyle{empty}

\begin{abstract}
  Deep Convolutional Neural Networks (CNNs) have become state-of-the art for computer vision and other signal processing tasks due to their superior accuracy. In recent years, large efforts have been made to reduce the computational costs of CNNs in order to achieve real-time operation on low-power embedded devices.
  
  Towards this goal we present BinArray, a custom hardware accelerator for CNNs with binary approximated weights. The binary approximation used in this paper is an improved version of a network compression technique initially suggested in \cite{Lin2017}. It drastically reduces the number of multiplications required per inference with no or very little accuracy degradation. \mbox{BinArray} easily scales and allows to compromise between hardware resource usage and throughput by means of three design parameters transparent to the user. Furthermore, it is possible to select between high accuracy or throughput dynamically during runtime. \mbox{BinArray} has been optimized at the register transfer level and operates at 400\,MHz as instruction-set processor within a heterogenous \mbox{XC7Z045-2} FPGA-SoC platform.
    
  Experimental results show that \mbox{BinArray} scales to match the performance of other accelerators like EdgeTPU \cite{EdgeTPU2019} for different network sizes. Even for the largest MobileNet only 50 \% of the target device and only 96 DSP blocks are utilized.
\end{abstract}


\section{Introduction}

Convolutional neural networks (CNN) have become state-of-the-art for machine vision and other signal processing tasks due to their superior classification accuracy. However, this superior accuracy is often accompanied by high computational complexity and memory intensity. This poses a challenge for the deployment of CNNs in all kind of embedded edge computing devices with limited resources and tight power constraints. In response to this challenge, the number of publications of CNN custom hardware accelerators has been growing over the past five years \cite{Sze2019}. Such custom designs are able to outperform general-purpose processors both with respect to throughput and energy efficiency.  

\subsection{Previous Work}

According to a recent survey \cite{Wang2019}, network approximations employed for custom hardware accelerators mainly fall into two categories: weight reduction and quantization. The second category can be further divided into fixed-point representation, logarithmic quantization, and binarization. Our accelerator design belongs to the latter sub-category, although by design not all fixed-point operations are eliminated. 

Full binarization of weights and activations as in BinaryNet \cite{Courbar2016} drastically reduces the complexity of inference operations, but also suffers from relatively poor accuracy performance with no means to control it. In contrast to full binarization, multi-level binary approximation of only weights replaces the majority of multiplications involved in a convolution with simple sign-changes, but retains fixed-point accumulation. This concept was first introduced by \cite{Rastegari2016} in the context of XNOR-Net, which uses one binary filter and some scaling factor to approximate the original filter weights. \cite{Lin2017} extended this concept with ABC-Net using a linear combination of several binary filters. This multi-level binarization achieved much better approximation of weight values and thus higher network accuracy. Unfortunately, the procedure developed in \cite{Lin2017} for finding appropriate binary filters and scaling factors for given network weights was not optimal and refined by the authors of \cite{Guo2017} shortly afterwards. Recently \cite{Zhu2020} then introduced piece wise approximation for binarizing weights and activations. However, they compared their results with the flawed procedure from \cite{Lin2017}, neglecting the results of \cite{Guo2017}.

In this paper we further improve the procedure from \cite{Guo2017} for finding an appropriate multi-level binary representation of weights. Based on the corresponding results, we argue that it is not necessary to binary encode weights {\em and} activations as done in \cite{Zhu2020} in order to achieve monotone accuracy increase. We then design a custom hardware accelerator for CNN inference using this approximation methodology. A key feature of this accelerator is that it can be easily scaled according to given accuracy and throughput requirements under given hardware resource constraints. To the best of our knowledge, this is the first time a hardware accelerator for CNNs approximated according to \cite{Lin2017, Guo2017} is proposed.

Closest to our work is ReBNet \cite{Ghasem2018}, which also provides a parameter to control the trade-off between throughput, accuracy and hardware resource usage. In contrast to our approach ReBNet uses single-level binary weights but multi-level binary activations. While we determine and train an optimal multi-level binary representation of weights offline, ReBNet binarizes weights and learns scaling factors for activations offline, and then performs multi-level binarization of activations during inference. This gives ReBNet the advantage of reduced memory footprint for network weights, but requires extra hardware resources for multi-level binarization of activations. In particular, \cite{Ghasem2018} reports a high usage of parallel multipliers (DSP blocks), which then even becomes the limiting hardware resource for two of their application examples (MNIST on XC7S50 and ImageNet on VCU108). With the binary approximation procedure used in this paper it is sufficient to only encode network weights and thus avoid the area and energy overhead associated with multi-level binarization of activations in hardware. 

\subsection{Contributions}

The contributions of this paper can be summarized as follows:
\begin{itemize}
	\item a method for multi-level binarization of CNN weights improving previous results from \cite{Lin2017, Guo2017} with respect to accuracy and monotone behavior.
	\item a novel systolic-array (SA) architecture for processing such binary approximated CNNs, which maximizes the reuse of features and thus reduces the required memory bandwidth. This SA processes all conventional CNN layers including max-pooling layers.
	\item an instruction-set accelerator for heterogeneous processing systems based on the proposed SA. This custom hardware accelerator, called \mbox{BinArray}, is entirely scalable in throughput, accuracy, and area by means of three parameters. 
	\item a register-transfer level implementation of \mbox{BinArray} running at \SI{400}{\mega\hertz} on a Xilinx Zynq XC7Z045-2 FPGA-SoC, which is faster than any previously reported designs for this platform.
\end{itemize}
In section~\ref{sec:binapprox}, the binary weight approximation methodology is explained. In section~\ref{sec:ops} we show how CNN layer operations are mapped to hardware with the architecture being developed bottom-up. In section~\ref{sec:proc}, the \mbox{BinArray} processing system of which the performance is evaluated in section~\ref{sec:results} is presented. The paper is concluded in section~\ref{sec:concl}.    
\section{Binary Approximated Weights}\label{sec:binapprox}

The proposed accelerator architecture is based on a multi-level binary representation of CNN weights as in \cite{Lin2017}. In this section this approximation is first formalized, followed by the presentation of an improved algorithm compared to \cite{Guo2017} for determining the coefficients.

\subsection{Approximation Formulation}\label{sub:binform}

\begin{figure}[t]
	\centering
	\includegraphics[width=\linewidth]{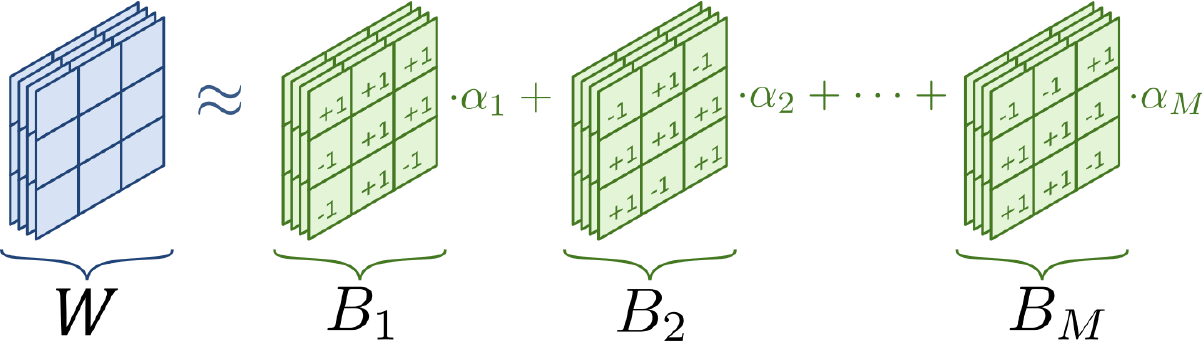}
	\caption{Binary Approximation of a real-valued filter kernel $W$ with $M$ binary filters $B_1,B_2,\ldots,B_M$.}
	\label{fig:bin_approx}
\end{figure}

As illustrated in Figure~\ref{fig:bin_approx}, the basic idea is to approximate a real-valued filter kernel $W$ by means of a linear combination of binary filter kernels $B_m$:
\begin{align}
	W\ \approx\ \sum_{m=1}^M B_m\cdot\alpha_m
	\label{eq:bin_approx_1}
\end{align}
with $B_m\in\mathbb{B}^3$ and $\mathbb{B} = \{+1,-1\}$. As explained in section~\ref{sec:ops}, this representation allows to drastically reduce the number of area/energy-costly  multiply-accumulate (MAC) operations during network inference, while preserving a mean to control network accuracy.

Each original filter coefficient $w_{i}$ is represented as a linear combination of the $i^\textrm{th}$ elements of the $M$ binary tensors $b_{i,m}\in\mathbb{B}$:
\begin{align}
	w_{i} \approx \sum_{m=1}^M b_{i,m} \cdot \alpha_m\ ,\qquad i=1,\ldots,N_c 
	\label{eq:bin_approx_2}
\end{align}
Thus, each coefficient $w_{i}$ of filter $W$ must be chosen from the same set $\omega$ of different values with $|\omega| = 2^M$.
\begin{align}
\begin{aligned}
\omega = &\{\alpha_1+\alpha_2+\ldots+\alpha_m,\\ 
&-\alpha_1 + \alpha_2 + \ldots + \alpha_m,\ \ldots\ , \\
&-\alpha_1-\alpha_2 - \ldots - \alpha_m\}
\end{aligned}
\end{align}
Approximation accuracy can then be controlled by the number $M$ of binary filters used, with larger $M$ yielding a more accurate approximation of the original filter $F$. 

\subsection{Approximation Procedure}\label{sub:binproc}

\subsubsection{Defining the Optimization Problem}
In order to approximate a given filter kernel $W$ a set $\boldsymbol{B}=\{B_1,\ldots,B_m\}$ of binary tensors and an associated set $\boldsymbol{\alpha}=\{\alpha_1,\ldots,\alpha_M\}$ of scaling factors must be computed. This can be formulated as a least-squares optimization problem, see \eqref{eq:optim}. Since it is not directly possible to optimize two parameters concurrently, either $\boldsymbol{B}$ or $\boldsymbol{\alpha}$ must be defined first. We follow \cite{Lin2017, Guo2017} and first determine the set of $M$ binary tensors, see section~\ref{subsub:binten}, and then obtain $\boldsymbol{\alpha}$ from solving
\begin{align}
\underset{\alpha}{\textrm{min }}J(\alpha)\ =\ \left\|W\ -\ \sum_{m=1}^M B_m\cdot\alpha_m\right\|^2
\label{eq:optim}
\end{align}
Flattening $W$ one can write \eqref{eq:bin_approx_2} as a set of linear equations
\begin{align}
\begin{bmatrix}
w_{1}\\
w_{2}\\
\vdots\\
w_{N_c}
\end{bmatrix} &\approx 
\begin{bmatrix}
b_{1,1} & \hdots & b_{1,M}\\
b_{2,1} & \hdots & b_{2,M}\\
\vdots & \ddots & \vdots\\
b_{N_c,1} & \hdots & b_{N_c,M} 
\end{bmatrix} \cdot
\begin{bmatrix}
\alpha_1\\
\vdots\\
\alpha_M
\end{bmatrix}
\label{eq:lin_eq}
\end{align}
which can be solved using a standard least-squares method to obtain the optimum $\boldsymbol{\alpha}$ for given $\boldsymbol{B}$. 
\subsubsection{Defining the Binary Tensors}\label{subsub:binten}
The original paper \cite{Lin2017} suggested to define the binary tensors $\boldsymbol{B}$ by splitting the weights into $M$ equidistant steps along the standard deviation of $W$. However, this is sub-optimal since it does not account for the symmetry of the linear combination in \eqref{eq:bin_approx_2} due to $b_{i,m}\in\mathbb{B}$. Therefore, the authors of \cite{Guo2017} suggested an improved procedure for determining $\boldsymbol{B}$, which is shown here as Algorithm~\ref{alg:1}.

\begin{figure}[t]
	\centering
	\includegraphics[width=\linewidth]{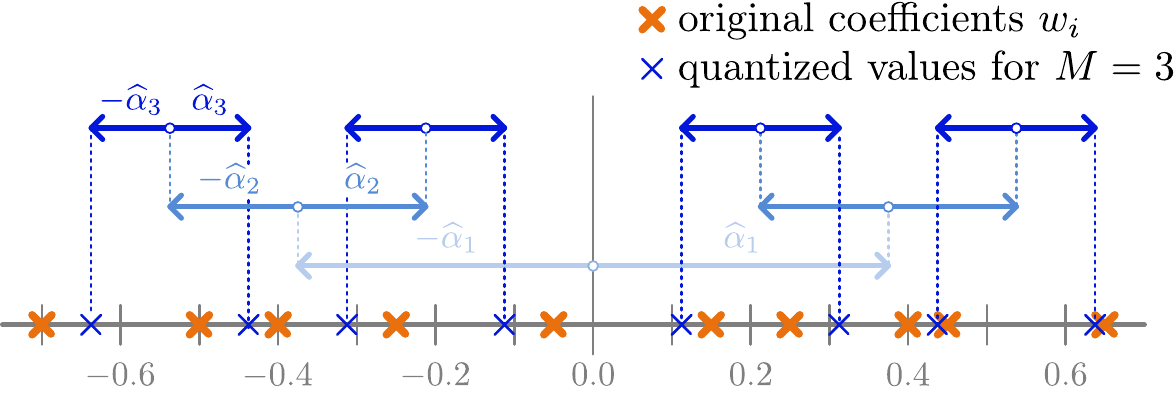}
	\caption{Concept of iteratively defining the binary tensors according to Algorithm~\ref{alg:1}. Shown are the first three iterations yielding $B_1, B_2, B_3$ and the associated $\hat{\alpha}_1,\hat{\alpha}_2,\hat{\alpha}_3$.}
	\label{fig:bintens_1}
\end{figure}

The rationale behind this procedure is as follows: For the first binary tensor the best approximation is $B_1 = \text{sign}(W)$ because we require $b_{i,m}\in\mathbb{B}$. Since the final scaling factor for $B_1$ is only available later from solving \eqref{eq:lin_eq}, the algorithm in step 4 estimates $\hat{\alpha}_1$ as the mean of the absolute value of all original filter coefficients. Step 5 then calculates the deviation of each filter coefficient from this estimate. These three steps are repeated to recursively obtain the desired number $M$ of binary tensors. Each subsequent tensor $B_m$ can be seen as an extension to the preceding tensor $B_{m-1}$, providing two times more weight values that can be represented and thus a better approximation of filter coefficients $w_i$, see Figure~\ref{fig:bintens_1}. Finally, the $M$ binary tensors are used to obtain the scaling factors $\boldsymbol{\alpha}$ from solving \eqref{eq:lin_eq}, in step 6. However, since only estimates  $\boldsymbol{\hat{\alpha}}$ were used to obtain the binary tensors  $\boldsymbol{B}$, wrong sign values may be assigned to individual elements $b_{i,m}$ resulting in larger than necessary approximation errors for filter weights. 

To avoid this, we suggest to recursively repeat the two-step procedure of running Algorithm~\ref{alg:1} to obtain  $\boldsymbol{B}$ and solving \eqref{eq:lin_eq} to get  $\boldsymbol{\alpha}$ until the optimum combination of binary tensors and scaling factors is found. This procedure is shown in Algorithm~\ref{alg:2}. The set of binary tensors and associated scaling factors is recursively updated, until no further improvement can be made and the binary tensors are stable. Since at this point the value of individual elements $b_{i,m}$ may start to oscillate between $+1$ and $-1$, the algorithm is aborted after a certain number of iterations $K$.

\subsection{Weight Compression}\label{sub:compr}

Besides reducing the number of MACs per inference, see section~\ref{sec:ops}, binary approximation as defined above also compresses the weights as long as $M$ is sufficiently small. Let $\textrm{bits}_w$ and $\textrm{bits}_\alpha$ be the number of bits used to represent the original filter coefficients $w_i$ and scaling factors  $\boldsymbol{\alpha}$, respectively. The compression factor achieved by binary approximation for a filter $W$ with $N_c$ elements plus one bias value is then given by
\begin{align}
	\textrm{compression factor}\ =\ \frac{(N_c+1)\cdot \textrm{bits}_w}{M\cdot(N_c+\textrm{bits}_\alpha)}\ \lessapprox\ \frac{\textrm{bits}_w}{M}
	\label{eq:cf}
\end{align}
with $N_c\gg \textrm{bits}_\alpha$ in most practical situations. Thus, assuming single-precision floating-point weights with $\textrm{bits}_w = 32$, compression factors will approach 16, 10.7, and 8 when using $M= 2, 3$ and 4, respectively. The coefficients of fully-connected layers can be approximated in the same way by using $M$ 1D binary tensors for each neuron.

Numerical results showing compression factors for real networks and comparing the accuracy achieved with Algorithm~\ref{alg:1} from \cite{Guo2017} and our enhanced Algorithm~\ref{alg:2} are given in section~\ref{ssub:acc}.

\begin{algorithm}[t] 
	\caption{Define a set $\boldsymbol{B}$ of $M$ binary tensors and then compute $\boldsymbol{\alpha}$ (according to \cite{Guo2017})} 
	\label{alg:1} 
	\DontPrintSemicolon
	$\Delta W \gets W$\;
	\For{$m=1$ \KwTo $M$}{
		$B_m \gets \text{sign}(\Delta W)$\;
		$\hat{\alpha}_m \gets \text{mean}(\Delta W\odot B_m)$\;
		$\Delta W \gets \Delta W - (B_m\cdot\hat{\alpha}_m$)\;
	}
	$\boldsymbol{\alpha} \gets $ solve \eqref{eq:lin_eq} with $\boldsymbol{B}$\;
\end{algorithm}

\begin{algorithm}[t] 
	\caption{Find sets $\boldsymbol{B}$ and $\boldsymbol{\alpha}$ of $M$ binary tensors and scaling factors recursively (our procedure)}
	\label{alg:2} 
	\DontPrintSemicolon
	$\boldsymbol{B},  \boldsymbol{\alpha} \gets $ Algorithm~\ref{alg:1}\;
	\texttt{iteration} = 0\;
	\Repeat{($\boldsymbol{B} = \boldsymbol{B}_{old}$) \texttt{or} (\texttt{iteration} = $K$)}{
		\texttt{iteration}$++$\;
		$\boldsymbol{B}_{old} \gets \boldsymbol{B}$ \;
		$\Delta W \gets W$\;
		\For{$m=1$ \KwTo $M$}{
			$B_m \gets \text{sign}(\Delta W)$\;
			$\Delta W \gets \Delta W - (B_m\cdot\alpha_m)$\;
		}
		$\boldsymbol{\alpha} \gets $ solve \eqref{eq:lin_eq} with $\boldsymbol{B}$\;}
\end{algorithm}

\section{Accelerated Operations}\label{sec:ops}
Conventional CNNs consist of a small set of layers that are processed in predetermined order. While the hyper-parameters vary across layers, the applied mathematical operations remain the same. This section describes the hardware implementation of all such operations supported by the \mbox{BinArray} accelerator. 

\subsection{Binary Dot Product}

The dot product is at the heart of convolution and dense layers operations. Let $\vec{x}$ be the vector of input activations and $\vec{w}$ the vector of weights, then the dot product $O$ is
\begin{align}
	O &= \sum_{i=1}^{N_c} x_i\cdot w_i
	\label{eq:dot_prod}
\end{align}
with $N_c$ being the number of coefficients. Employing \eqref{eq:bin_approx_2} the binary dot product can be written as
\begin{align}
	O &\approx \sum_{m=1}^M \alpha_m \sum_{i=1}^{N_c} x_i \cdot b_{i,m}\ .
	\label{eq:bin_dot_prod}
\end{align}
Since  $b_{i,m}\in\mathbb{B}$, the number of real-valued multiplications is reduced from $N_c$ to $M$. As demonstrated in section~\ref{sec:results}, $M\ll N_c$ can be achieved for any practical filter size used in CNNs with no or little accuracy loss. In order to parallelize the computation of multiple dot products while maximizing input feature reuse, we follow the design paradigm of systolic arrays in a bottom-up way. 

\begin{figure}[t]
	\centering
	\includegraphics[width = \linewidth]{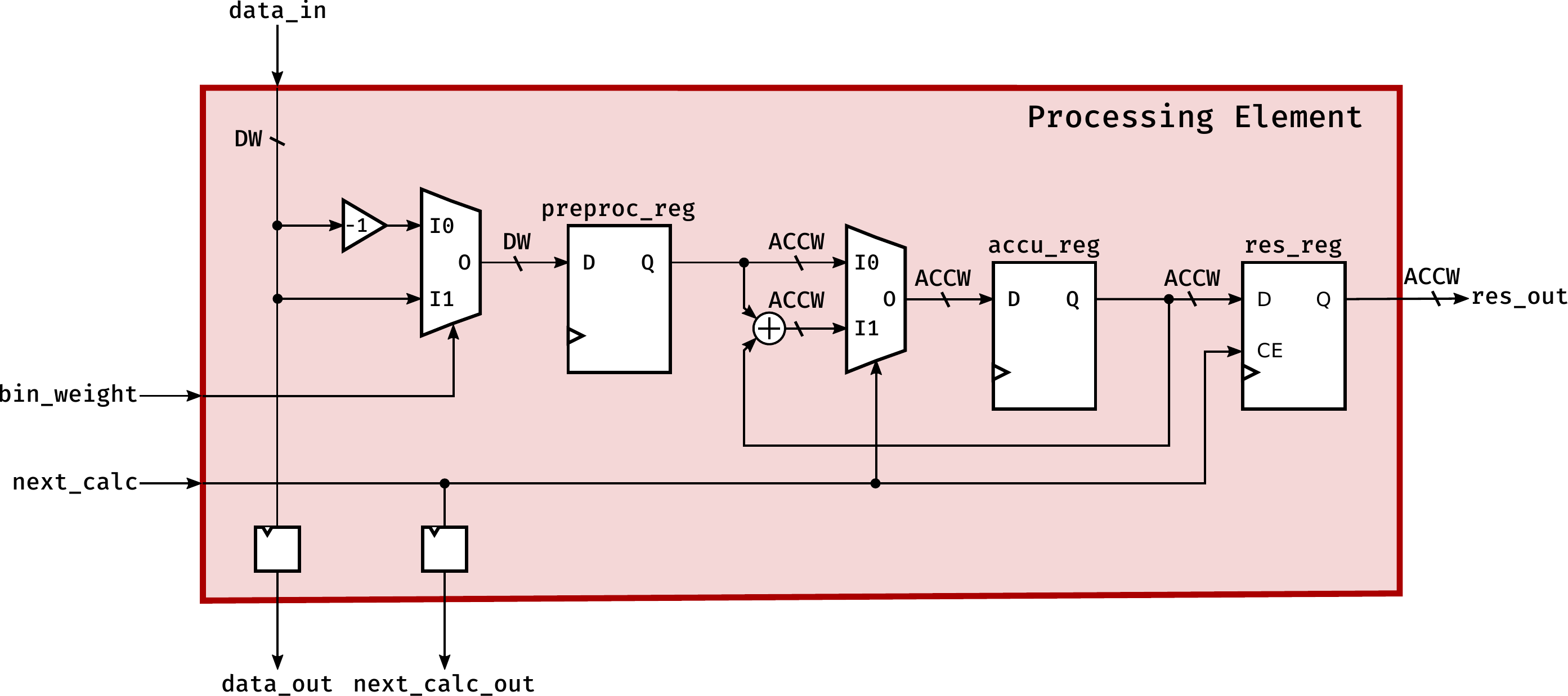}
	\caption{RTL diagram of the processing element (PE) consisting of a conditional sign change, a single adder and an accumulation register. The output equals the partial result of a binary dot product according to \eqref{eq:part_result}.}
	\label{fig:pe}
\end{figure}

The key building block of the array are processing elements (PE) as shown in Figure~\ref{fig:pe}. In every clock cycle (cc), each PE can take an input activation $x_i$, calculate its additive inverse according to the corresponding binary weight $b_{i,m}$ and add this value to an accumulation register. Thus, the PE computes
\begin{align}
	p_m = \sum_{i=1}^{N_c} b_{i,m} \cdot x_i
	\label{eq:part_result}
\end{align}
where $p_m$ is the partial result of the dot product of the $m^{\text{th}}$ binary filter, see \eqref{eq:bin_dot_prod}. 
Once the computation of $p_m$ is completed, the result is shifted into the PE output register for further processing by downstream logic, the accumulator cleared, and the next partial result calculation started without idle cycles.

\begin{figure}[t]
	\centering
	\includegraphics[width=0.9\linewidth]{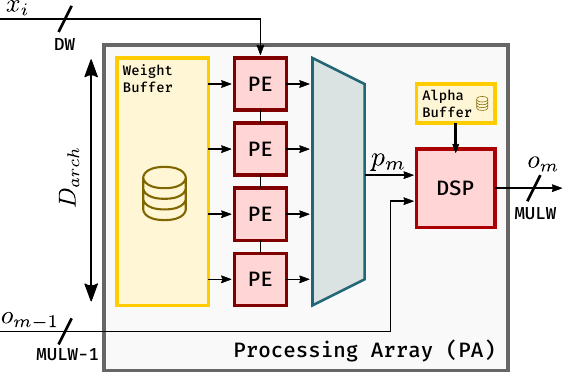}
	\caption{The PA with inputs $x_i$ for input features and $o_{m-1}$ for the result of a previous PA. The PA stores the 1-Bit weights for the operation described in \eqref{eq:matrix_vect} in a local weight buffer. Scaling values $\alpha_m$ for $D_{arch}$ channels are stored in a local $\alpha$ buffer.}
	\label{fig:pa}
\end{figure}

Several PEs are then vertically connected to form a processing array (PA) as shown in Figure~\ref{fig:pa}. Each PE receives the input feature from the preceding PE and forwards it to the next PE with one cc delay, thus facilitating both input feature reuse and time-sharing of costly DSP hardware blocks. The number of PEs per PA and thus the number of channels that can be computed in parallel is denoted $D_{arch}$, the first configurable design parameter of the \mbox{BinArray} accelerator.

\begin{figure}[t]
	\centering
	\includegraphics[width=\linewidth]{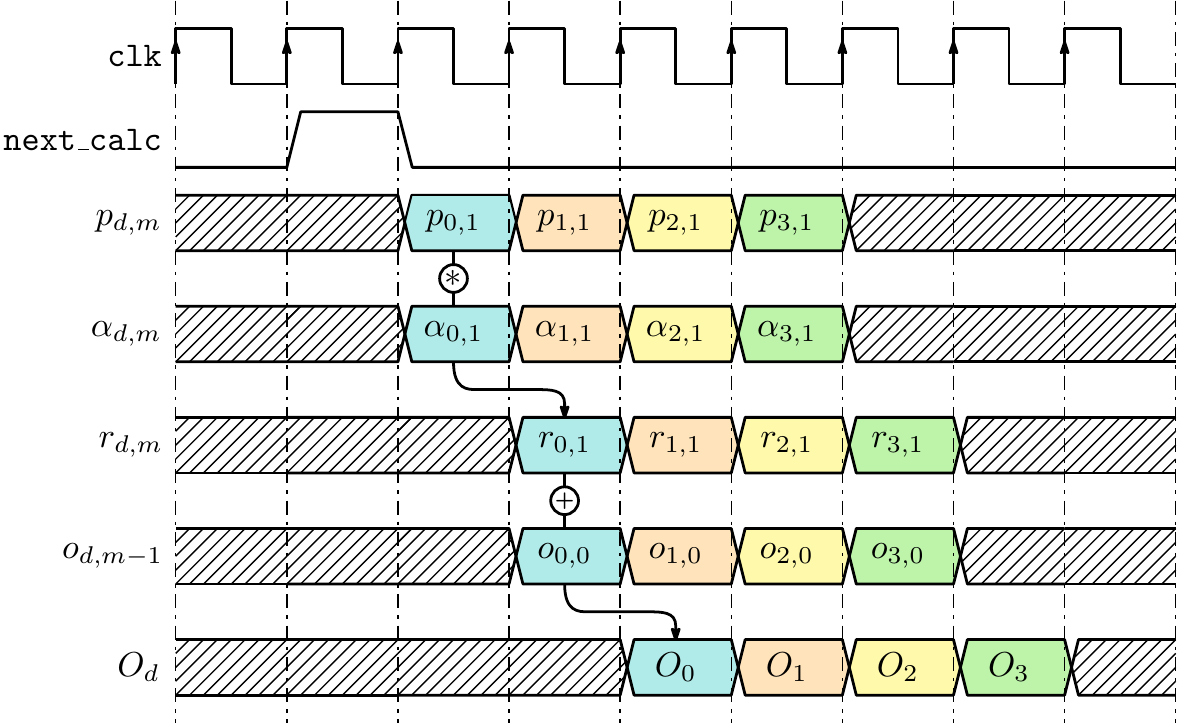}
	\caption{Timing (without pipeline registers) of binary dot product calculation in an PA for $D_{arch}=4$ and $M=2$. Signal \texttt{next\_calc} triggers the serialized output of the partial sums $p_{d,m}$ from the $D_{arch}$ PEs, which are then multiplied with the corresponding $\alpha_{d,m}$. The resulting product $r_{d,m}$ is added to the output of the previous PA, leading to the final results $O_d$.}
	\label{fig:PA_output}
\end{figure}

Mathematically, a PA performs the following matrix-vector multiplication for input vector $\vec{x}$ and one binary weight matrix $B_m$:
\begin{align}
	\underset{\scriptscriptstyle D_{arch} \times 1}{\vec{p_m}} &= \underset{\scriptscriptstyle D_{arch} \times N_c}{B_m} \times \underset{\scriptscriptstyle N_c\times 1}{\vec{x}}
	\label{eq:matrix_vect}
\end{align}

Since $B_m \in \mathbb{B}^2$, $D_{arch}$ output channels require $N_c \cdot D_{arch}$ bits of storage. In order to keep the weights close to the PEs, a dual-port BRAM is used as storage for weights. 

The one cc delay in input activation forwarding results in a staggered output stream of all $D_{arch}$ channels as illustrated in Figure~\ref{fig:PA_output} for $D_{arch} = 4$. Also shown is the recursive computation of the final dot product $O_d$ from the $M$ partial products $r_{d,m}$ of this channel, i.e.
\begin{align}
	o_m = \underbrace{p_{d,m} \cdot \alpha_{d,m}}_{r_{d,m}} + o_{d,m-1}, \quad m\in [0,M-1]
	\label{eq:out_rec}
\end{align}
where $O_d=o_{d,M-1}$. The multiply-add operations required for calculating the output $o_{d,m}$ for all $D_{arch}$ channels of one PA can be time-multiplexed using a single DSP macro. The required $\alpha$'s are stored in a small distributed-RAM memory in fixed-point format.

While it is possible to compute \eqref{eq:out_rec} sequentially on one PA,
even better reuse of input features results when multiple PAs operate in parallel on the $M$ different binary filters according to \eqref{eq:bin_dot_prod}. This is established by grouping a fixed number of $M_{arch}$ PAs in parrallel, such that outputs $o_{d,m}$ are cascaded as shown in Figure~\ref{fig:PA_output}. Since the outputs are in fixed-point format, we align partial results with a configurable barrel shifter. The first PA associated with $m=0$ takes in the bias $\beta_d$ of the output channel.  

The second configurable design parameter $M_{arch}$ represents the hardware-supported $M$ and controls inference accuracy and throughput. We will return to this in section~\ref{sec:proc}.

\subsection{Activation Function and Pooling}

Conventional CNNs use activation functions at the output of convolutional layers directly followed by pooling layers. While activation functions keep subsequent CNN layers linearly independent, pooling reduces the locality of the information. \mbox{BinArray} combines activation function and pooling into a common operation for efficiency reasons.

\begin{figure}[t]
	\centering
	\includegraphics[width=0.9\linewidth]{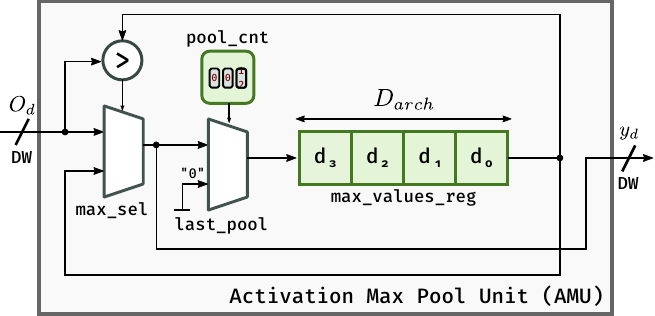}
	\caption{The Activation Max-Pooling Unit (AMU) receives its input from the last PA and performs ReLU activation and max-pooling. The shift register holds the intermediate maximum values of $D_{arch}$ output samples according to \eqref{eq:mpool}. This is required as data from the PAs is in \textsl{channel-first} order.}
	\label{fig:amu}
\end{figure}

Since almost all current CNN architectures use some form of rectification function, \mbox{BinArray} implements \mbox{ReLU} as activation function:
\begin{align}
	\label{eq:relu}
	\textsl{ReLU}(x) &= \textsl{max}(x,0)\ .
\end{align} 
Pooling layers reduce tensor dimensions by means of either downsampling (input shape is an integer multiple of the output shape) or resampling (input shape is not an integer multiple of the output shape). While resampling requires extra calculations and adds data dependencies, downsampling can be directly integrated into the output data stream. Therefore, \mbox{BinArray} implements max-pooling layers with downsampling only.

\mbox{ReLU} activation and max-pooling are jointly implemented by the activation and max-pooling unit (AMU) using the commutative property of these two operations as shown in Figure~\ref{fig:amu}. First, max-pooling is performed as
\begin{align}
 \label{eq:mpool}
 y_{k+1} = \textsl{max}(y_{k}, O_{d,k}), \quad k\in [0,N_p-1]
\end{align} 
where $O_{d,k}$ is the $k^{\text{th}}$ sample of the binary dot product from the PA and $N_p$ is the downsampling factor. With $y_0 = 0$, a positive $y_{N_p}$ results if and only if at least one $O_k$ was positive, which corresponds to ReLU as in \eqref{eq:relu}. 

Since the PA outputs are arranged channel-wise, see Figure~\ref{fig:PA_output}, but downsampling is done depth-wise, the intermediate maximum values for $D_{arch}$ output channels must be stored in the AMU. This is done with a shift-register of length $D_{arch}$ which is updated based on maximum-comparison with new input values. During the $N_p^{\text{th}}$ iteration the AMU will output the $D_{arch}$ maximum values of the pooling window and the shift register is reset to zero, ensuring correct ReLU activation for the next pooling window. 

\subsection{Fixed-Point Quantization}

All input and output activations are represented in fixed-point format with \texttt{DW=8} bits. However, output data $O_d$ of the DSP blocks within PAs are represented with \texttt{MULW=28} bits, see Figure~\ref{fig:pa}. The reason being that accumulation within the cascade of DSP blocks is done with full precision. Eventually, the result of the last PA is quantized back to \texttt{DW} bits before being sent to the AMU. This quantization is relative to a predefined, layer-dependent binary point position, rounding off LSBs and saturating in case of overflow. Within the AMU data word width does not change, see Figure~\ref{fig:amu}.

\section{BinArray Processing System}\label{sec:proc}

This section first shows how the low-level processing blocks introduced in section~\ref{sec:ops} form a systolic array. Then the infrastructure required to operate one or more of such arrays by means of an instruction-set based processor within a heterogeneous processing platform is described.

\subsection{Systolic Array}

\begin{figure}[t]
	\centering
	\includegraphics[width=\linewidth]{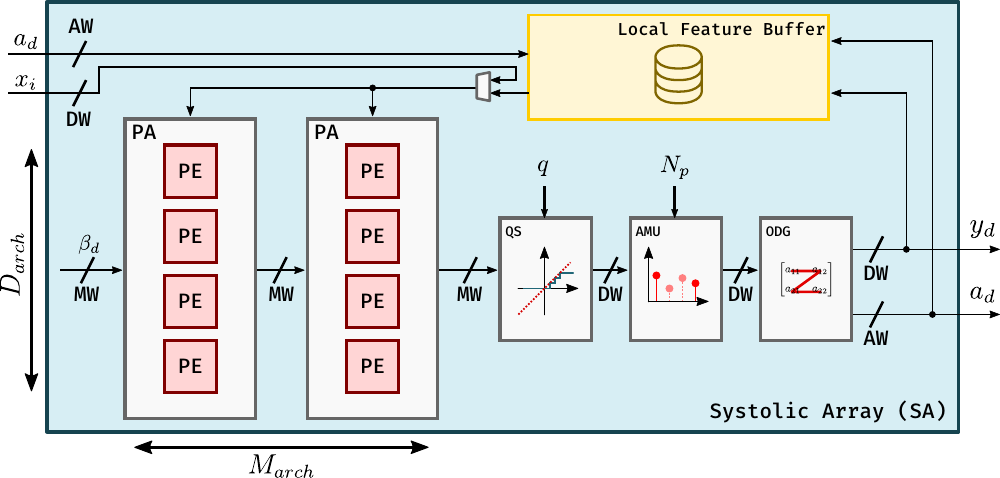}
	\caption{The Systolic Array (SA) can compute the output of both convolution and dense layers with binary approximated weights autonomously. The SA also performs the activation of output neurons as well as max-pooling.}
	\label{fig:sa}
\end{figure}

The components introduced in section~\ref{sec:ops} are combined to form a systolic array (SA) as shown in Figure~\ref{fig:sa}. The $M_{arch}$ PAs arranged horizontally each consist of $D_{arch}$ PEs distributed vertically. Such SA can compute $M_{arch}$ binary filters on $D_{arch}$ output channels in parallel. PEs in the same row of the array operate on the same output channel, while PEs in the same column process the same binary filter $m$. 

The array output values are then quantized by the QS block to the supported data width \texttt{DW} before being sent to the AMU for downsampling. Furthermore, the SA also contains a local feature buffer to hold all input and output features involved in the current computation. These can either be the input and output features of an entire layer, or some tile of it. The local feature buffer is implemented as dual-port RAM, such that input feature reading and output feature writing can happen simultaneously. This allows hidden layers of the CNN to be processed back-to-back without global data communication, provided they entirely fit into the local feature buffer. 

The buffer is organized in row-major order. The required write addresses are provided by the output data gatherer (ODG). This block assigns a row-major address to the output values which arrive from the AMU in channel-first order.

\subsection{Feature Buffer Address Generator}

The address generator unit (AGU) is responsible for accessing input features in the order required for processing within the SA. Address generation depends on the layer type, since the same feature is used multiple times in convolutional layers but only once in dense layers.

\subsubsection{Convolutional Layers}
\begin{figure}[t]
	\centering
	\includegraphics[width=\linewidth]{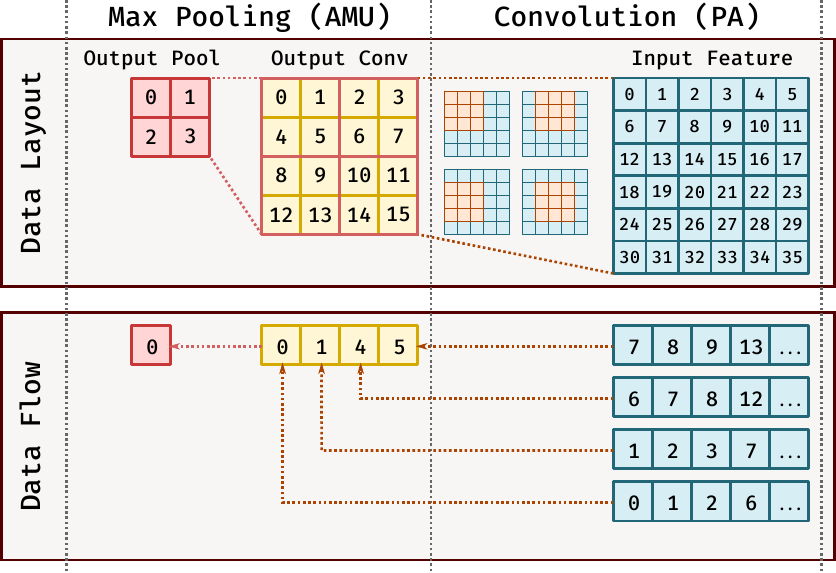}
	\caption{Top: Data layout with convolution windows needed to produce the first four pooling outputs. Bottom: Required data flow in SA for the first pooling output.}
	\label{fig:dataflow}
\end{figure}

The kernel window is slid across the input feature to generate a two dimensional output. Traditionally, this sliding is performed row-wise, i.e.\ an imaginary anchor point is first shifted along the same row before it is shifted down to the next row. However, for the proposed SA this is not applicable, as downsampling is carried out by the AMU in the output data stream directly. This requires the anchor points of consecutive convolutions to fall within the pooling window currently being processed. 
The corresponding processing order is illustrated in Figure~\ref{fig:dataflow} for a $3\times 3$ convolution window and a $2\times 2$ pooling window. In the top part the data layout shows the convolution windows required to produce the first four pooling outputs. The data flow in the bottom part of Figure~\ref{fig:dataflow} shows the order in which input features must be processed to compute the first pooling output.
In order to calculate input feature addresses without multiplications the AGU uses Algorithm~\ref{alg:3} to maintain the following anchor points (absolute addresses) and indexes (address offsets):
\newline
\begin{tabular}{lcl}
	$a_{cv}$ & : & starting point of current convolution window \\
	$a_{po}$ & : & starting point of current pooling window \\
	$a_{cl}$ & : & first index of current row in current pooling window \\
	$i_{cl}$ & : & first column index of current input window \\
	$p_w$    & : & current column index within pooling window \\
	$p_h$    & : & current row index within pooling window \\
\end{tabular}
\newline
\begin{algorithm}[t] 
	\caption{Anchor point calculation for conv layers}
	\label{alg:3} 
	\DontPrintSemicolon
	\KwData{$\ i_{cl}, p_w, p_h, a_{cv}, a_{po}, a_{cl}\gets 0$}
	\BlankLine
	\uIf(\tcc*[f]{move conv. to next column}){$p_w < W_P-1$}
	{$a_{cv} \gets a_{cv} + 1$\;
		$p_w \gets p_w + 1$\;}
	\uElseIf(\tcc*[f]{move conv. to next row}){$p_h < H_p-1$}
	{
		$a_{cv},\,a_{cl} \gets a_{cl} + W_I$\;
		$p_h \gets p_h + 1$\;
		$p_w \gets 0$\;}
	\uElseIf(\tcc*[f]{move pool $\rightarrow$}){$i_{cl} < W_I - W_B - W_P + 1$}
	{$a_{cv}, a_{cl}, a_{po} \gets a_{po} + W_P$\;
		$i_{cl} \gets i_{cl} + W_P$\;
		$p_w, p_h \gets 0$\;}
	\Else(\tcc*[f]{move pool $\downarrow$})
	{$a_{cv}, a_{cl} \gets a_{cv} + W_B$\;
		$a_{po} \gets a_{cv} + W_B + W_P$\;
		$p_w, p_h, i_{cl} \gets 0$\;}
\end{algorithm} 

\begin{figure}[t]
	\centering
	\includegraphics[width=0.99\linewidth]{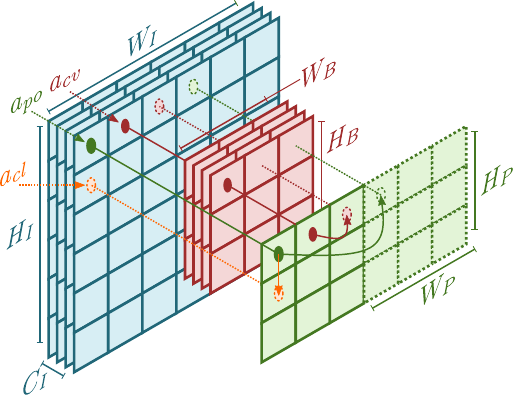}
	\caption{Convolution over a feature (blue) using a filter (red) and subsequent pooling (green) with $W,H,C$ denoting the corresponding dimensions. Additional anchors calculated according to Algorithm~\ref{alg:3} determine the next convolution window.}
	\label{fig:addresses}
\end{figure}

Given convolution anchor $a_{cv}$, it is straightforward to generate the absolute addresses of the input features required for the current convolution. 
To obtain convolution anchor $a_{cv}$ Algorithm ~\ref{alg:3} is required to test four cases. The first two cases move convolution anchor $a_{cv}$ within the current pooling window only, using indexes $p_w$ and $p_h$. When all convolutions within the current pooling window are completed, the pooling window is shifted horizontally. This is repeated until the pooling window reaches the horizontal limit of the input feature. Signaled by index $i_{cl}$, the pooling window is then shifted downwards once this limit is hit. This process continues until the complete input feature window has been processed. Note that whenever pooling anchor $a_{po}$ is moved, the convolution anchor is set to the first address of the new pooling window.

Figure~\ref{fig:addresses} visualizes this process with current and next anchors shown as filled and dashed circles, respectively. In the example shown, the next convolution starts directly adjacent to the current column and the pooling anchor $a_{po}$ is shifted to the left by one pooling unit $W_P$. Furthermore, $a_{cl}$ stores the address of the first pixel in the current row of the current pooling window. This index is necessary for computing the convolution anchor for the next row.

\subsubsection{Dense Layers}
Since dense layers do not require pooling, the AMU is bypassed for this layer type. In this case the AGU implements a simple linear counter.

\subsection{Control Unit}

The control unit has been designed to let one or more systolic arrays perform CNN inference autonomously. More specifically, the control unit allows to operate \mbox{BinArray} as instruction-set processor within a heterogeneous processing platform consisting of a software-based CPU system and programmable logic, as for example the Xilinx Zynq FPGA-SoC. While the CPU handles input and output data at the application level, \mbox{BinArray} can process a complete CNN from input to output layer without further interaction with the CPU.

The control unit supports a small set of 32-bit instructions. By means of these instructions, the user programs the network to be processed. An example of such CNN processing program is shown in Listing 1. Note that these programs can be easily generated by a compiler from any high-level network specification. The CPU then loads the program to an instruction memory in programmable logic from where the control unit reads and executes the program.

The control unit features a set of configuration registers that hold the parameters for the SAs and infrastructure blocks to process a given layer. These registers are written by the \texttt{STI} instruction. Once a layer has been processed, the control unit re-configures the parameters according to the CNN processing program and starts processing the subsequent layer.

The \texttt{HLT} instruction pauses execution of the processing program until a trigger is received from the CPU. This allows to synchronize the loading and unloading of new input images and results by the CPU with layer processing in programmable logic. It could also be used to process inner network layers with operations not supported by \mbox{BinArray} in software. The \texttt{CONV} instruction stalls program execution until processing of the current layer is completed. This could both be a convolutional or dense layer. Finally, the unconditional branching instruction \texttt{BRA} at the end of the processing program jumps back to the beginning of the program once inference for one image has been completed. 

Although possible, the CU design does not pipeline the execution of instructions. The rational behind this being that the number of cc for the setup of an entire layer (STI instructions) are negligible compared to the number of cc required for processing the layer.

\lstdefinestyle{customasm}{
	xleftmargin=\parindent,
	language=[x86masm]Assembler,
	basicstyle=\footnotesize\ttfamily,
	commentstyle=\scriptsize,
	numbers=left,
	columns=c[flexible],
	xleftmargin=16pt
}
\lstset{escapechar=@,style=customasm}
\begin{lstlisting}[float,caption=CNN Processing Programm for BinArray,label=lst1:mxm]
STI r0 W_I=48   ; Set input width to 48 pixels
STI r1 W_B=7    ; Set kernel width to 7 pixels
@$\cdots$@
HLT             ; Wait for trigger from PS
@\textbf{CONV}@ 0	    ; Start CONV of 1st layer
STI r0 W_I=21   ; Set input width to 21 pixels 
STI r1 W_B=4    ; Set kernel width to 4 pixels
@$\cdots$@
@\textbf{CONV}@ 1	    ; 2nd CONV layer, mark last layer
@\textbf{BRA}@ 1	     ; Branch to step 1
\end{lstlisting}

\subsection{BinArray System}

\begin{figure}[t]
	\centering
	\includegraphics[width=\linewidth]{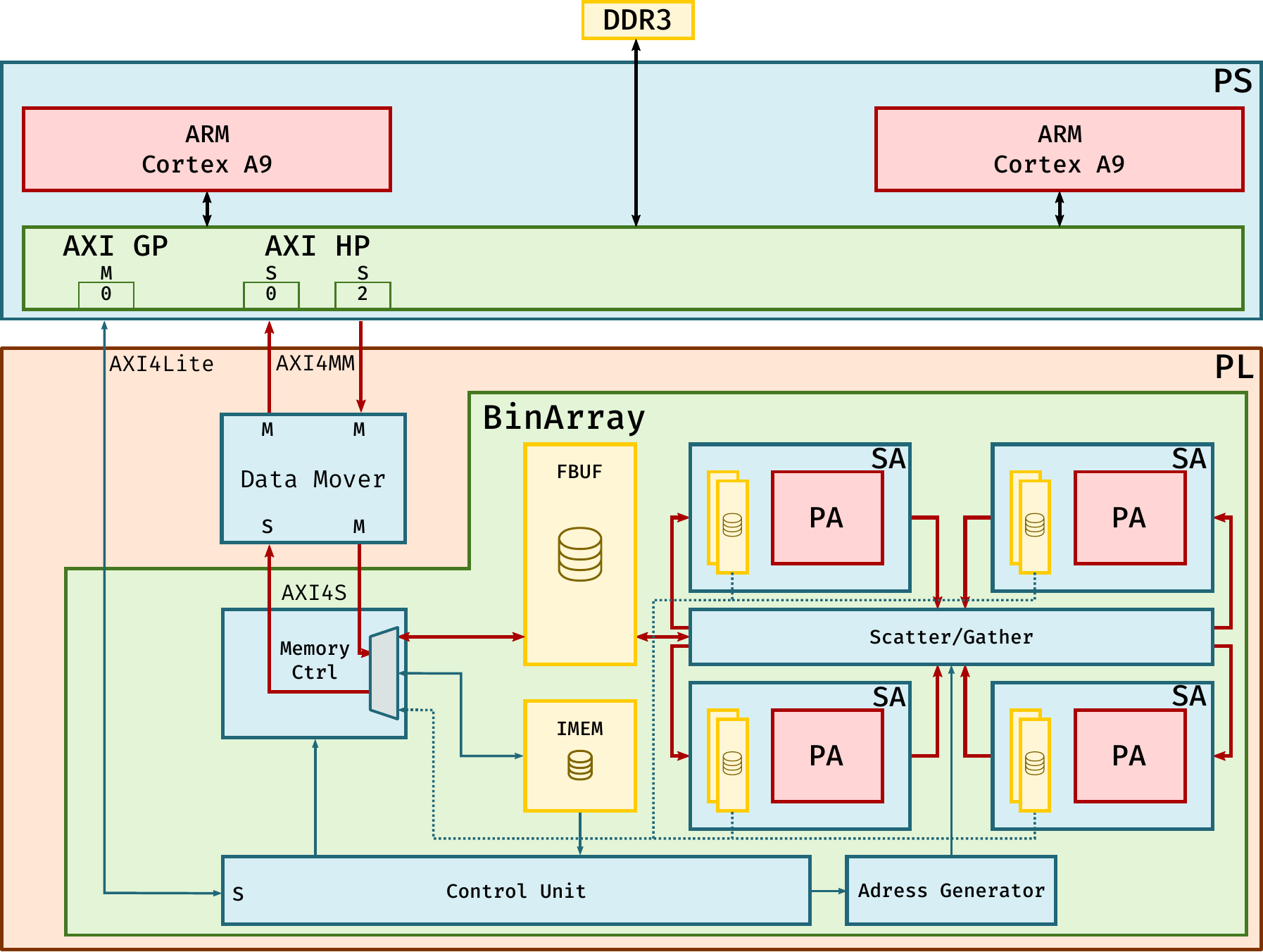}
	\caption{The BinArray System. A single  SA is responsible for the calculation of the BACNN. Additionally, there is a global FBUF and a CU with its own IMEM. Data transmission of features and weights happens through DMA with a data mover.}
	\label{fig:sys}
\end{figure}

Figure~\ref{fig:sys} shows the \mbox{BinArray} accelerator in the programmable logic part of the FPGA-SoC connected by two high performance (HP) and one general purpose (GP)  AXI interface to the CPU system. The GP port connects basic registers of BinArray, providing the option to enable and disable the accelerator. The HP ports transport the features from external DDR3 memory to the global feature buffer (FBUF) by means of a DMA block. The FBUF is implemented as ping-pong buffer, allowing to pipeline data acquisition in the CPU system and CNN inference by BinArray. Additionally, this data channel allows to compute unsupported intermediate network layers in the CPU and transfer the results back to the accelerator for processing of the remaining layers.

Multiple SA can be instantiated and work on tiles of the same input feature in parallel. The number of arrays $N_{SA}$ is the third configurable design parameter of \mbox{BinArray}. If $N_{SA}>1$, the data flow between FBUF and the different arrays is controlled by a scatter/gather block as indicated in Figure~\ref{fig:sys}. 

Table~\ref{tab:despar} summarizes all three design parameters and their meaning. For example, a configuration	with $N_{SA}=1$, $D_{arch}=16$, and $M_{arch}=2$ will be referred to as \mbox{BinArray}[1, 16, 2].

Note the relation between binary approximation parameter $M$, see section~\ref{sec:binapprox}, and the hardware design parameter $M_{arch}$. If, for example, some application requires $M=4$ to achieve the desired inference accuracy, but the hardware architecture was designed to only process $M_{arch}=2$ binary tensors in parallel. Then, two passes per convolution could be performed for high accuracy, while only one pass is done for high throughput. Hence, \mbox{BinArray} allows to choose between a high-accuracy and high-throughput mode on the same hardware at run time.%
\begin{table}[b]
	\caption{Configurable design parameters of \mbox{BinArray}.}
	\label{tab:despar}
		\centering
	\setlength\extrarowheight{2pt}
	\begin{tabular}{lll}
		\hline
		Parameter & Meaning & Effect\\
		\hline
		$N_{SA}$   & \# of parallel SA & throughput\\
		$D_{arch}$ & \# of output channels & throughput\\
		$M_{arch}$ & \# of binary tensors & throughput/accuracy\\
		\hline
	\end{tabular}%
\end{table}%
\subsection{Performance Model} \label{sec:perf_model}
For throughput estimation of \mbox{BinArray} an analytical model has been developed based on the following paradigms: 
\begin{enumerate}
	\item Each PE can perform one accumulation per cc. Multiplications happen in parallel with the accumulation of input feature, see Figure~\ref{fig:PA_output}. In other words, the remaining multiplications with scaling factors $\alpha_m$ do not affect throughput, only latency. 
	\item Tiling of input features is only performed in the width and height dimensions, but not in the depth dimension. This makes convolutions atomic, which eliminates further post-processing. 
	\item The SA pipeline is not stalled for loading input features, because features either fit in the local buffer or can be loaded without delay from the global buffer. 
\end{enumerate}
First, the number of output features to be computed per layer is
\begin{align}
\label{dimO}
\dim(O) &= \{U, V, D\}\\
&=  \{\frac{W_I-W_B+2P}{S}+1,\frac{H_I-H_P+2P}{S}+1,D\}\nonumber.
\end{align}
where $W_I,H_I$ and $W_B,H_B$ are width and height of the input and kernel respectively. $P$ denotes padding of the input and $S$ the filter stride. 
Each SA calculates $M_\text{arch}$ binary filters in parallel. Since we aim to support both $M$ (high-throughput mode) and $2M$ (high-accuracy mode) on the same hardware, the effective number of logical SAs (LSA) is
\begin{equation}
\label{eq:nlas}
N_\text{LAS} = N_\text{SA} /\lceil\frac{M}{M_\text{arch}}\rceil.
\end{equation}
Note that choosing $M < M_\text{arch}$ does not result in a faster computation, but leads to idle SA columns. 

Second, multiple output channels $D$ are also calculated in parallel. $D_\text{max}$ is the maximum number of output channels that can be calculated concurrently with the accelerator. If the number of output channels $D$ is smaller than the total number of rows of all SAs in the accelerator, we apply tiling of the input in order to keep the PEs busy. The number of tiles equals
\begin{equation}
\label{eq:NT}
\{N_{T} =\lfloor N_\text{LSA}/ \lceil\frac{D}{D_\text{arch}}\rceil \rfloor \quad | \frac{W_I}{N_T}>1 \land \frac{H_I}{N_T}>1\}\ .
\end{equation}
Here $\lceil\frac{D}{D_\text{arch}}\rceil$ assures that no further tiling can be done if the number of filters is smaller than the number of PEs in a SA. In this case, the remaining PEs will be idle. If, however, the number of output channels is larger than $D_\text{max}$, multiple BinArray passes are needed to compute the layers output. This is given by
\begin{equation}
\label{eq:Npass}
N_\text{pass} = \lceil\max(1,\frac{D}{D_\text{arch} \cdot N_\text{LSA}})\rceil.
\end{equation}
Combining these relations, the number of cc required to compute the output features for one layer is 
\begin{equation}
N_\text{cc} = \frac{W_I\cdot H_I\cdot C_I\cdot W_B\cdot H_I\cdot N_\text{pass}}{N_T}\ .
\label{eq:tput}
\end{equation}
Note that the degree of hardware parallelism expressed by $D_\text{arch}$ and $M_\text{arch}$ enters \eqref{eq:tput} via $N_T$ and $N_\text{pass}$ in a non-straightforward way.

\section{Results}\label{sec:results}

In this section, the setup and models used for all experiments are described. Then, results and interpretations for network accuracy, clock speed, throughput and resource usage are provided.

\subsection{Experimental Setup}

\subsubsection{Networks and Data Sets}\label{ssub:ref}
The following two combinations of reference networks and data sets were used to evaluate both the binary weight approximation scheme described in section~\ref{sec:binapprox} and the \mbox{BinArray} processor implementation and its performance.
\begin{itemize}
	\item CNN-A\\ A smaller CNN with a total of 9M MACs over two convolutional layers ($5@7\times7\times3$, $150@4\times4\times5$) and three dense layers (1350 $\to$ 340 $\to$ 490 $\to$ 43 neurons) used on the GTSRB data set. \cite{Stallkamp2012}
	\item CNN-B1\\ MobileNetV1 with $\rho=0.57, \alpha=0.5$ \cite{Howard2017} and a total of 49M MACs trained on ImageNet with an input size of $128\times 128$.
	\item CNN-B2\\ MobileNetV1 with $\rho=1, \alpha=1$ \cite{Howard2017} and a total of 569M MACs trained on ImageNet with an input size of $224\times 224$.
\end{itemize}  
The depth-wise layers of MobileNetV1 were approximated channel-wise, as there exists only a single convolution filter.

\subsubsection{Hardware Implementation}
\begin{figure}[t]
	\centering
	\includegraphics[width=\linewidth]{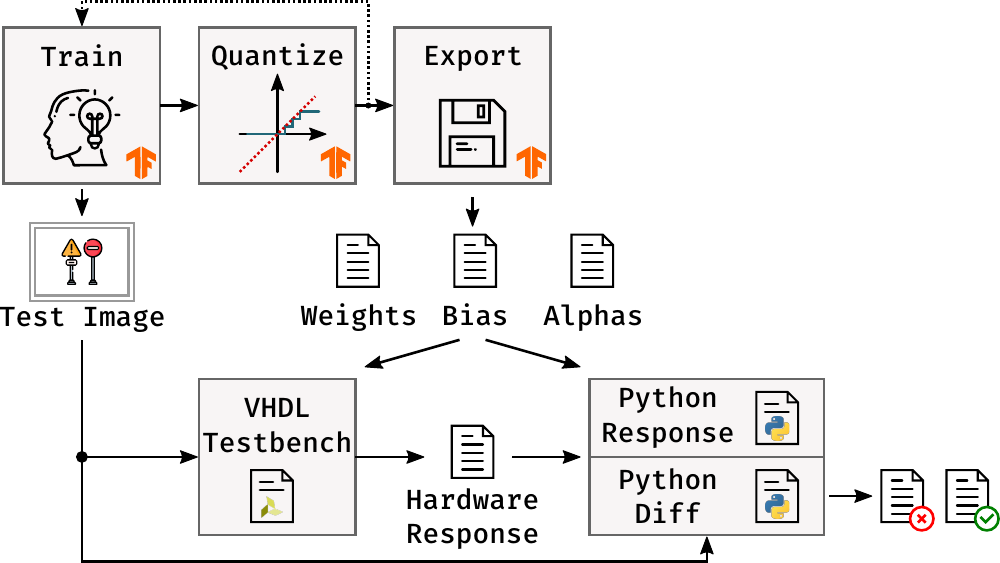}
	\caption{Verification setup for the VHDL implementation of \mbox{BinArray}.}
	\label{fig:test}
\end{figure}

The \mbox{BinArray} processor system for $N_{SA}=1$ has been implemented in VHDL and verified for bit-accurate operation using the setup shown in Figure~\ref{fig:test}. The trained weights were exported from TensorFlow together with some sample images. The VHDL simulation response for these images were then compared to the results of a bit-accurate Python model.

\subsubsection{Performance Estimation}
Throughput is calculated by means of the analytical performance model developed in section~\ref{sec:perf_model}. 

To verify this model, a cycle-accurate simulation of processing the first two layers of Network 1 was performed using the VHDL test bench environment shown in Figure~\ref{fig:test}. The analytical model \eqref{eq:tput} predicts a processing time of 466'668 cc for these two layers, while VHDL simulation required 467'200 cc to complete. The discrepancy is due to the pipelined implementation of the systolic array SA and the instruction processing time of the control unit, which both are not accounted for by the analytical model. However, the resulting error of \SI{-1.1}{\permille} is sufficiently small to be neglected.

Reference networks \mbox{CNN-B1/B2} include depth-wise convolution layers for which \mbox{BinArray} has not yet been optimized. These layers can still be processed, using only a single PE per PA, however. Thus, for depth-wise convolution layers $D_\text{arch}=1$ is assumed in \eqref{eq:Npass}, eliminating output channel parallelism for these layers.  

\subsection{Experimental Results}

\subsubsection{Network Accuracy}\label{ssub:acc}
Table~\ref{tab:binapprox} shows compression factors and compares the network accuracy achieved with and without retraining using Algorithm~\ref{alg:1} from \cite{Guo2017} and our enhanced Algorithm~\ref{alg:2}. Retraining was done for one epoch using the straight-through estimation proposed by \cite{Courbar2016} for gradient calculation. Algorithm~\ref{alg:2} was run for $K=100$ iterations. In all cases, binary approximation provides useful initial values for retraining. To prevent the optimizer from unlearning this valuable starting point, a lower learning rate is mandatory. CNN-A was retrained with the Adam optimizer with $\alpha = 1\times 10^{-4}, \beta_1 = 0.9, \beta_2 = 0.999$. On CNN-B, Adam was susceptible for exploding gradients, which was solved by switching to SGD with a momentum of $\beta = 0.9$. The learning rate $\alpha$ is initialized with $\alpha_0 = 5\times 10^{-4}$ and decayed exponentially over retraining.

The compression factor was calculated with $\si{bits}_\alpha=8$ and $\si{bits}_w=32$. As can be seen, the compression factor as a function of $M$ approaches the predicted values according to \eqref{eq:cf} for all networks. With retraining, reasonable accuracies are achieved even for small $M$. With larger $M$, accuracy degradation becomes negligible in most cases. Compared to a quantized \textsl{int8} implementation, binary approximation improves the compression factor by at least $20\%$ even for $M=6$. Algorithm~\ref{alg:2} outperforms Algorithm~\ref{alg:1} in almost every situation. In particular, the desired monotone increase in accuracy with increasing $M$ is only achieved by Algorithm~\ref{alg:2}, both with and without retraining. We therefore attribute the lack of monotony reported in \cite{Zhu2020} to their use of the flawed approximation procedure from \cite{Lin2017}.

Note that while the results in Table~\ref{tab:binapprox} are obtained by using the same $M$ for all layers in the network, the \mbox{BinArray} accelerator can deal with individual $M$ for each layer. This can be useful for layers, which do not benefit from additional accuracy, like most dense layers for classification at the end of the network.

\begin{table}
	\caption{Compression factor (cf) and Top-1 accuracies with two different binary approximation procedures as function of $M$. For each network the single-precision floating-point accuracy is indicated for comparison.}
	\label{tab:binapprox}
		\centering
	\setlength\extrarowheight{2pt}
	\begin{tabular}{crcccc}
		\hline
		&    & \multicolumn{2}{c}{acc. w/ Algorithm~\ref{alg:1}} & \multicolumn{2}{c}{acc. w/ Algorithm~\ref{alg:2}} \\
		$M$ & \multicolumn{1}{c}{cf} & no retrain & w/ retrain & no retrain & w/ retrain\\
		\hline
		\multicolumn{6}{c}{CNN-A (baseline acc.\ 97.86\%)} \\
		\hline
		2	& 15.8 & 84.68\%	&97.09\%	& 87.43\%	& 97.13\%\\
		3	& 10.6 & 93.40\%	&97.51\% 	& 95.92\%	& 97.29\%\\
		4	&  7.9 & 95.64\%	&96.60\% 	& 97.51\%	& 98.01\%\\
		\hline
		\multicolumn{6}{c}{CNN-B1 (baseline acc. 56.3\%)} \\
		\hline
		4	&  7.6 & 0.10\% & 43.17\% & 0.18\% & 51.55\% \\
		5	&  6.1 & 0.08\% & 46.29\% & 0.64\% & 54.46\% \\
		6	&  5.1 & 0.10\% & 50.96\% & 5.03\% & 55.03\% \\
		\hline
		\multicolumn{6}{c}{CNN-B2 (baseline acc.\ 70.9\%)} \\
	    \hline
		4	&  7.9  & 0.11\% & 46.90\% &  0.2\% & 47.82\% \\
		5	&  6.2 & 0.12\% & 46.84\% & 6.8\%& 53.59\% \\
		6	&  5.2  & 0.08\% & 51.23\% & 25.2\% & 69.10\%  \\
		\hline
	\end{tabular}

\end{table}

\subsubsection{Clock Speed}\label{ssub:clk}
For the target 28\,nm ZYNQ device \mbox{XC7Z045-2} our VHDL implementation achieved timing closure at 400\,MHz clock frequency using \cite{Xilinx2020}. Experiments showed that on the more recent 16\,nm \mbox{UltraScale+} technology a clock frequency of up to 666\,MHz is feasible. Compared to e.g.\ \cite{Ghasem2018}, who reported 200\,MHz on 20\,nm \mbox{UltraScale} technology with high-level synthesis, this significantly higher clock speed is a result of the register-transfer-level implementation of \mbox{BinArray} together with careful gate-level optimizations at critical points of the data path. This confirms the general rule that bit-level algorithms, like CNNs with binary encoding techniques, are not well suited for high-level synthesis when targeting high-speed and/or area-efficient implementations. 

\subsubsection{Throughput}
\begin{table*}
	\caption{Throughput in frames per second (FPS) of a hypothetical \SI{1}{\GOPS} CPU and several configurations of \mbox{BinArray}.\newline [1,32,2] means $N_{SA}=1$, $D_{arch}=32$, $M_{arch}=2$ }
	\label{tab:perf}
	\centering
	\setlength\extrarowheight{2pt}
	\begin{tabular}{llrrrrrrrrr}
		\hline
		CNN  & $M$ & \multicolumn{4}{c}{BinArray}                 & CPU   & Edge$\quad$            &  Eyeriss \\
		     &&  [1,8,2] & [1,32,2] & [4,32,4] & [16,32,4] &       & TPU \cite{EdgeTPU2019} & V2 \cite{Sze2017}\\
		\hline
		-A   & 2 & 354.2 &  819.8 &     -   &  -       & 111.8  & -     &  -   \\
		-B1  & 4 & 46.7  &  92.5  &  728.4 & 3845.5  &  20.6 & -     &  1282.1    \\
	    -B2  & 4 & 2.6   &  7.7   &  74.3 & 350.0    &  1.8  & 416.7 &  -    \\
	   	-B1  & 6 & 20.0  &  55.7  &  364.2 & 1036.0  &  20.6 & -     &  1282.1    \\
	    -B2  & 6 & 1.8   &  5.8   &  37.1  &  175.0   & 1.8  &  416.7 &  -   \\
		\hline
	\end{tabular}
\end{table*}

In Table~\ref{tab:perf} the performance of different \mbox{BinArray} configurations for the reference networks/data sets defined in section~\ref{ssub:ref} are compared to a hypothetical processing unit with \SI{1}{\GOPS}. This, for instance, could be a CPU running at \SI{1}{\giga\hertz} with a single MAC unit being constantly utilized, or, a SIMD processor with equivalent throughput. For the throughput of this hypothetical CPU only the MAC operations of all network layers are taken into account. All other operations (ReLU, max-pooling) are neglected. 

\mbox{BinArray} throughput figures are based on the analytical model \eqref{eq:tput} with a clock frequency of 400\,MHz, see section~\ref{ssub:clk}. For the small network CNN-A for instance, configuration \mbox{BinArray[1, 32, 2]} can be used, which employs only one SA, i.e.\ $N_{SA}=1$. Furthermore, setting $M_{arch}=2$ provides for switching between a high-throughput mode with $M=2$ and a high-accuracy mode with $M=4$ at runtime. Note, however, that in this case accuracy degradation would be marginal even with $M=2$, see Table~\ref{tab:binapprox}. 

For MobileNetV1, the parameters in the final dense layer are responsible for nearly half of the total number of parameters of the network. Although those parameters would fit into the on-chip BRAM of the target platform, they are only needed for less than 1\% of the total processing time. It thus makes sense for a heterogeneous system like the Xilinx Zynq FPGA SoC to offload this final task to the CPU. Not only does the CPU have access to sufficient storage for those parameters, it can also process the global average pooling before the final layer. Furthermore, calculating the average in hardware would  require additional area just for this minor task.
Thanks to the user configurable parameters of BinArray, a wide range of networks can be accelerated. Depending on applications constraints, a selection of throughput, accuracy and resource utilization is possible. For larger CNNs like CNN-B1\& B2, BinArray can even reach the performance of larger accelerators in ASIC technology by having $N_{SA}>1$.

The results show that an increase of $D_{arch}$ only results in higher throughput if the processed layer has equal or more channels than $D_{arch}$. This can be best observed in CNN-A, where a $4\times$ increase in $D_{arch}$ only results in a $2\times$ increase in throughput. The source of this non linear increase lies in the first layer of CNN-A. Here, in the case of $D_{arch}=32$, just 15\% of PEs can be utilized.

\subsubsection{Area \& Energy Efficiency}
\begin{table}
	\caption{Resource utilization of target ZYNQ device XC7Z045 for different \mbox{BinArray} configurations in \%. [1,32,2] means $N_{SA}=1$, $D_{arch}=32$, $M_{arch}=2$}
	\label{tab:area}
	\centering
	\setlength\extrarowheight{2pt}
	\begin{tabular}{lrrrrr}
		\hline
	                &         & \multicolumn{4}{c}{BinArray} \\
		            &  Total  & [1,8,2] & [1,32,2] & [4,32,4] & [16,32,4] \\
		\hline
		LUT   		&  218,600  &   0.78 & 1.68      & 13.32  &  52.74    \\
		FF    		&  437,200  &   0.53 & 1.22      & 8.11   &  32.01     \\
		BRAM CNN-A	&   \SI{19.2}{\mega\bit}  &   1.15 & 1.15      & 6.19   &  24.2   \\
		BRAM CNN-B	&    \SI{19.2}{\mega\bit}  &   23.72    & 23.94   & 28.85   &  46.90   \\
		DSP  		&    900    &   0.22    &  0.22    &  1.78   &  7.11  \\
		\hline
	\end{tabular}
\end{table}

Table~\ref{tab:area} shows the FPGA resource usage for the same \mbox{BinArray} configurations used for performance evaluation in Table~\ref{tab:perf}. Numbers for $N_{SA}>1$ are estimated based on utilization figures for $N_{SA}=1$. Based on current estimations, an overhead of 200 FF and 230 LUTs per SA was added. 

Note that the number of DSP blocks will always equal \mbox{$N_{SA}\cdot M_{arch}$} since exactly one MAC operation is used per PA. While the weights in BinArray are multi-level binary encoded, the activations are encoded in regular 2's complement fixed-point format. This eliminates the need for additional encoding/decoding circuitry, which was the case in \cite{Ghasem2018}. BinArray on the other hand drastically reduces the number of DSP slices. These valuable resources are thus available to other applications in a heterogeneous compute system. 

For storing activations and weights, BRAMs provide fast access to memory for BinArray. While for CNN-A all parameters fit into the weight buffer BRAMs, CNN-B1/B2 need additional global buffers to store the parameters of the convolution layers.
According to \cite{sze2020}, energy cost of a  32-bit read access is around $100\times$ lower for internal SRAM than off-chip SDRAM. Consequently, for large CNNs like CNN-B1/B2 a global 4Mb BRAM weight buffer is instantiated in the device. Interestingly, about the same energy ratio exists between an 8-bit addition and a 32-bit multiplication (both for fixed- and floating-point types). Thus, assuming only external data access and 32-bit multiplications for CPU operation, both memory and arithmetic energy consumption would be $100\times$ less with BinArray. Using a factor of ten as safety margin, we hence conservatively assume that CNN inference on BinArray can be performed at least $10\times$ more energy efficient than with a hypothetical CPU implemented in the same technology.         
\section{Conclusion}\label{sec:concl}

This paper showed that multi-level binary representation of weights is an effective network approximation technique based on which efficient custom hardware accelerators can be designed. More specifically, we presented a scalable solution that allows to control the compromise between accuracy and throughput under given hardware resource constraints by means of three design parameters. An analytical throughput model depending on these parameters has been developed. This model, together with known hardware resource usage, facilitates the use of our accelerator in an end-to-end framework similar to the one suggested in \cite{Ghasem2018}. Such framework can automatically generate CNN accelerators optimized for the application requirements at hand. Unlike with \cite{Ghasem2018}, the degree of parallelism for our accelerator is not limited by costly DSP blocks.

\bibliography{main}
\bibliographystyle{IEEEbib}

\end{document}